\documentclass{article}[12pt]
\usepackage{axodraw}
\newcommand{\e} {\mbox{\rm e}} 
\def\Clpt{\hbox{Cl}_2\left(\frac{\pi}{3}\right)} 

\newcommand{\ord}{\mathcal{O}}
\def\draftdate{\relax}
\def\mpar#1{\relax}
\def\mua{\relax}
\def\mda{\relax}
\def\mla{\relax}
\def\draft{
\def\thtystars{******************************}
\def\sixtystars{\thtystars\thtystars}
\typeout{}
\typeout{\sixtystars**}
\typeout{* Draft mode!
         For final version remove \protect\draft\space in source file *}
\typeout{\sixtystars**}
\typeout{}
\def\draftdate{\today}
\def\mua{\marginpar[\boldmath\hfil$\uparrow$]%
                   {\boldmath$\uparrow$\hfil}%
                    \typeout{marginpar: $\uparrow$}\ignorespaces}
\def\mda{\marginpar[\boldmath\hfil$\downarrow$]%
                   {\boldmath$\downarrow$\hfil}%
                    \typeout{marginpar: $\downarrow$}\ignorespaces}
\def\mla{\marginpar[\boldmath\hfil$\rightarrow$]%
                   {\boldmath$\leftarrow $\hfil}%
                    \typeout{marginpar: $\leftrightarrow$}\ignorespaces}
\def\Mua{\marginpar[\boldmath\hfil$\Uparrow$]%
                   {\boldmath$\Uparrow$\hfil}%
                    \typeout{marginpar: $\Uparrow$}\ignorespaces}
\def\Mda{\marginpar[\boldmath\hfil$\Downarrow$]%
                   {\boldmath$\Downarrow$\hfil}%
                    \typeout{marginpar: $\Downarrow$}\ignorespaces}
\def\Mla{\marginpar[\boldmath\hfil$\Rightarrow$]%
                   {\boldmath$\Leftarrow $\hfil}%
                    \typeout{marginpar: $\Leftrightarrow$}\ignorespaces}
\def\mpar##1{\marginpar{\hbadness10000%
                      \sloppy\hfuzz10pt\boldmath\bf##1}%
                      \typeout{marginpar: ##1}\ignorespaces}

\overfullrule 5pt
\oddsidemargin -15mm
\marginparwidth 29mm
}

\begin{document}
\unitlength1cm
\begin{titlepage}
\renewcommand{\thefootnote}{\fnsymbol{footnote}}
\vspace*{-1cm}
\draftdate
\begin{flushright}
TTP05-06\\
SFB/CPP-05-14\\
hep-ph/0505041\\
Mai 2005
\end{flushright}
\vskip 3.5cm
\begin{center}
{\Large\bf Precise numerical evaluation of the two loop sunrise graph 
Master Integrals in the equal mass case. } 
\vskip 1.cm
{\large  S.~Pozzorini} and {\large E.~Remiddi}\footnote{Supported by
    Alexander-von-Humboldt Stiftung \\ 
    permanent address: Dipartimento di Fisica,
    Universit\`{a} di Bologna, I-40126 Bologna, Italy \\ 
    e-mail: {\tt ettore.remiddi@bo.infn.it } } 
\vskip .7cm
{\it Institut f\"ur Theoretische Teilchenphysik,
Universit\"at Karlsruhe, D-76128 Karlsruhe, Germany}
\end{center}
\vskip 2.6cm

\begin{abstract} 
We present a double precision routine in Fortran for the precise and fast 
numerical evaluation of the two Master Integrals (MIs) of the 
equal mass two-loop sunrise graph for arbitrary momentum transfer in 
$d=2$ and $d=4$ dimensions. The routine implements the accelerated 
power series expansions obtained by solving the corresponding 
differential equations for the MIs at their singular points. 
With a maximum of 22 terms for the worst case expansion 
a relative precision of better than a part in $10^{15}$ is achieved 
for arbitrary real values of the momentum transfer. 
\end{abstract}
\vfill
\end{titlepage}
\newpage

{\bf\large PROGRAM SUMMARY}
\vspace{4mm}
\begin{sloppypar}
\noindent   
{\em Title of program\/}: {\tt sunem} \\[2mm]
{\em Version\/}: 1.0 \\[2mm]
{\em Release\/}: 1  \\[2mm]
{\em Catalogue identifier\/}: \\[2mm]
{\em Program obtainable from\/}: {\tt http://www-ttp.physik.uni-karlsruhe.de/Progdata/%
} \\[2mm]
{\em Computers\/}: all \\[2mm]
{\em Operating system\/}: all \\[2mm]
{\em Program language\/}: {\tt FORTRAN77} \\[2mm]
{\em Memory required to execute\/}: Size: 1532k \\[2mm]
{\em No.\ of bits in a word\/}: up to 32 \\[2mm]
{\em No.\ of processors used\/}: 1 \\[2mm]
{\em No.\ of bytes in distributed program\/}: 34589 \\[2mm]
{\em Distribution format\/}: ASCII \\[2mm]
{\em Other programs called\/}: none \\[2mm]
{\em External files needed\/}: none \\[2mm]
{\em Nature of the physical problem\/}: numerical evaluation of the 
two Master Integrals of the equal mass two-loop sunrise Feynman graph 
for arbitrary momentum transfer in $d=2$ and $d=4$ dimensions. \\[2mm]
{\em Method of solution\/}: accelerated power series expansions 
obtained by solving the  differential equations for the MIs 
at their singular points. 
With a maximum of 22 terms for the worse case expansion a relative precision 
of better than a part in $10^{15}$ is achieved for arbitrary real values 
of the momentum transfer. \\[2mm] 
{\em Restrictions on complexity of the problem\/}: 
limited to real momentum transfer and equal internal masses. \\[2mm]
{\em Typical running time\/}: approximately 1 $\mu$s to evaluate the four Master integrals for a fixed momentum transfer value on a Pentium IV/3 GHz Linux PC. 
\end{sloppypar}

\newpage

{\bf\large LONG WRITE-UP}

\section{ Introduction. } 
The continuous progress in the evaluation of radiative correections 
to various elementary particle processes, driven by the increasing 
number of current and future precise experimental results, is faced 
with the problem of the fast and precise numerical evaluation of those 
radiative corrections that
cannot be expressed in terms of 
known analytic functions for which numerical routines are already 
available. 
\par 
In this paper we discuss a {\tt FORTRAN} routine for the precise and fast 
numerical evaluation of the two Master Integrals (MIs) associated 
with the two-loop selfmass sunrise graph for equal masses, based on 
the results of \cite{Laporta:2004rb}, where their 
analytic properties were established by studying the linear system of 
first order inhomogeneous differential equations satisfied by the MIs 
themselves. \par
As a first obvious step, 
the linear system can be rewritten as a second order 
(inhomogeneous) linear differential equation for one of the MIs (from now 
on referred to as the main MI, or simply the MI), while the other MI can 
be expressed in terms of the first and its derivative. 
The accurate investigation of the second order equation shows that the 
immaginary part of the MI (which satisfies the associated homogeneous 
equation) is an ellyptic function of complicated argument (a fact 
known however since a long time), while the real part is likely to 
belong to a new family of analytic functions, which could not be 
traced back to the usual Nielsen's polylogarithms nor to their 
generalizations \cite{Remiddi:1999ew}. \par 
Nevertheless, the differential equation can be used to obtain a 
convenient integral representation of the solution and, among other 
things, the explicit limiting behaviour of the MI under investigation 
at zero momentum transfer, at the pseudothreshold, at the threshold 
and at asymptotic momenta. Given those limiting values as starting 
points, the differential equation can then be used to generate the 
coefficients of the relevant expansions up to any required order. 
\par 
In this paper we show that those expansions can be used for writing 
a (double precision) {\tt FORTRAN} routine for the fast and precise 
numerical evaluation of both the MIs, within the $d$-continuous 
regularization scheme, in $d=2$ and $d=4$ dimensions. 
According to \cite{Laporta:2004rb}, it is indeed 
sufficient to work out the relevant expansions only for the main MI 
in $d=2$ dimensions by direct iterative solution of the differential 
equation; the other MI in $d=2$ dimensions, as well as both MIs 
in $d=4$ dimensions, are then expressed in terms of the first MI 
in  $d=2$ dimensions and its derivatives (needless to say, 
differentiation becomes a 
simple algebraic operation when dealing with expansions). 
\par 
In the next 3 Sections we discuss the relevant expansions of the 
first MI in $d=2$ dimensions for physical momentum around zero, 
at threshold and at infinity. In Section 6 we describe how to use 
the results of the previous Section for implementing the {\tt FORTRAN} 
routine for the evaluation of the two MIs at $d=2$ and $d=4$ 
dimensions. 
In order to reach a relative precision of better than $1\times 10^{-15}$
(which is essentialy the limit of double precision {\tt FORTRAN})
for any value of the 
squared momentun transfer $s$ in the whole real axis, 
$-\infty < s < +\infty$, it is sufficient to keep a maximum of 
22 terms in the various expansions. 

\section{ Definition and notation. } 
\begin{figure}\begin{center}
\setlength{\unitlength}{1pt}
\begin{picture}(120.,104.)(-60,-52.)
\Line(-80,0.)(-40,0.)
\Vertex(-40,0.){2}
\Line(-40,0.)(40,0)
\Vertex(40,0.){2}
\Line(40,0.)(80,0)
\CArc(0,0)(40,0,360)
\end{picture}
\end{center}
\caption{The 2-loop sunrise graph.}
\label{fig:diagram}
\end{figure}
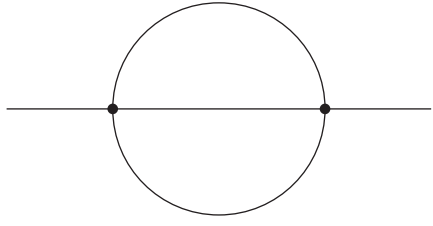

The two MIs associated with the (equal mass) 2-loop sunrise
graph depicted in Fig.~\ref{fig:diagram}
are defined, in the $d$-continuous regularization scheme, as 
\newcommand{\intm}{\mathcal{D}^d}
\begin{eqnarray} 
      S(d,z) &=& \int \frac{\intm k_1 \, \intm k_2 } 
                 {(k_1^2+1)(k_2^2+1)[(p-k_1-k_2)^2+1]},\nonumber\\ 
    S_1(d,z) &=& \int \frac{\intm k_1 \, \intm k_2 } 
                 {(k_1^2+1)^2(k_2^2+1)[(p-k_1-k_2)^2+1]},
\label{defMIs} 
\end{eqnarray} 
where
\begin{equation}
        \intm k=\frac{1}{\Gamma(3-\frac{d}{2})} 
                \frac{\mathrm{d}^d k}{4\pi^{\frac{d}{2}}},
\end{equation}
$p$ is the external momentum, 
$ p^2 = z = - s $ (when $p$ is Euclidean or spacelike, $z$ is positive 
and $s$ is negative). 
One has, for arbitrary $d$ 
\begin{equation} 
  S_1(d,z) = \frac{1}{3}\left[ - (d-3) + z\frac{d}{dz} \right] S(d,z) \ . 
\label{S1} 
\end{equation} 
At $d=2$ $S(d,z)$ and $S_1(d,z)$ are both finite, 
\begin{eqnarray} 
      S(d,z)   &=& S(2,z) + \ord(d-2) \ , \nonumber\\ 
      S_1(d,z) &=& S_1(2,z) + \ord(d-2) \ , 
\label{Sat2} 
\end{eqnarray} 
and from now on we will simply consider $S(2,z), S_1(2,z)$. 
$S(2,z)$ satisfies the second order inhomogenous equation 
\begin{equation}
  \biggl\{ \ \ \frac{d^2}{dz^2} +
         \left[\frac{1}{z} + \frac{1}{z+1} + \frac{1}{z+9}
        \right] \frac{d}{dz}
      + \left[ \frac{1}{3z} - \frac{1}{4(z+1)} - \frac{1}{12(z+9)}
         \right] \biggr\} S(2,z) = \frac{3}{8z(z+1)(z+9)} \ , 
\label{diffeq} 
\end{equation} 
which will be used to derive all the required expansions. 
\par 
Around $d=4$ $S(d,z)$ develops a double pole in $(d-4)$, so that it can be 
written as 
\begin{equation} 
   S(d,z) = \frac{1}{(d-4)^2} S^{(-2)}(4,z) + \frac{1}{d-4} S^{(-1)}(4,z) 
          + S^{(0)}(4,z) +\ord(d-4) \ .
\label{expS} 
\end{equation} 
From \cite{Laporta:2004rb} one has 
\begin{eqnarray} 
  S^{(-2)}(4,z) &=& - \frac{3}{8} \ , \nonumber\\ 
  S^{(-1)}(4,z) &=& \frac{1}{32}(z+18) \ , \nonumber\\ 
   S^{(0)}(4,z) &=& \frac{1}{12}(z+1)(z+9) \left( 
                    1 + (z-3)\frac{d}{dz} \right) S(2,z) \nonumber\\ 
                  && - \frac{1}{128}(72+13z) \ . 
\label{S(0,4)} 
\end{eqnarray} 
From Eq.(\ref{S1}), recalling that $S_1(2,z)$ is finite while 
$S_1(d,z)$ can be expanded around $d=4$ as $S(d,z)$ Eq.(\ref{expS}),
we find 
\begin{eqnarray} 
  S_1(2,z) &=& \frac{1}{3}\left[ 1 + z\frac{d}{dz} \right] S(2,z) \ , 
                             \label{S1(2)} \\ 
  S_1^{(-2)}(4,z) &=& \frac{1}{8} \ , \nonumber\\ 
  S_1^{(-1)}(4,z) &=& - \frac{1}{16} \ , \nonumber\\ 
   S_1^{(0)}(4,z) &=& \frac{1}{3} \left[  - 1 + z\frac{d}{dz} \right] 
                      S^{(0)}(4,z) - \frac{1}{96}(z+18) \ . 
\label{S1(0,4)} 
\end{eqnarray} 
We will work out explicitly, from the differential equations, 
the expansions of $S(2,z)$ in the various regions; we will then 
use Eq.s(\ref{S1(2)},\ref{S(0,4)},\ref{S1(0,4)}) for obtaining the 
expansions of $S_1(2,z), S^{(0)}(4,z)$ and $S_1^{(0)}(4,z)$ 
in terms of the corresponding expansions of $S(2,z)$ .

\section{ The expansion around $z=0$. } 
The point $z=0$ (or $s=0$) is a singular point of the 
differential equation Eq.(\ref{diffeq}); the investigation carried 
out in \cite{Laporta:2004rb} shows 
that the most general solution $G(z)$ consists of a regular part 
and a singular part behaving like $\ln{z}$ at $z=0$, so that its 
expansion is 
\begin{equation} 
     G(z) = \sum_{n=0}^\infty a_n z^n 
          + \ln{z} \sum_{n=0}^\infty b_n z^n \ . 
\label{Gexpat0} 
\end{equation} 
Once the expansion Eq.(\ref{Gexpat0}) is substituted in Eq.(\ref{diffeq}), 
the differential equation takes the form of a system of linear algebraic 
equations for the two sets of coefficients $a_n$, $b_n$, and can be solved 
recursively expressing the coefficient of order $n$ in terms of the 
coefficients of lower order. The procedure requires two initial conditions, 
{\it i.e.} the two coefficients $a_0$ and $b_0$. 
\par 
As discussed in \cite{Laporta:2004rb} in the case of $S(2,z)$ 
the two initial conditions are 
\begin{eqnarray} 
   a_0 &=& \frac{\sqrt{3}}{12}\Clpt = 0.146494...\ , \nonumber\\ 
   b_0 &=& 0 \ , 
\label{z0init} 
\end{eqnarray} 
where $\mbox{\rm Cl}_2\left(\frac{\pi}{3}\right)$ is the Clausen function 
of argument $\pi/3$, while the vanishing of $b_0$ is due to the 
regularity of $S(2,z)$ at $z=0$. 
\par 
Therefore, the expansion for $S(2,z)$ reads simply 
\begin{equation} 
     S(2,z) = \sum_{n=0}^\infty a_n z^n \ , 
\label{expat0} 
\end{equation} 
and on account of the explicit value of $a_0$ provided by Eq.(\ref{z0init}) 
the differential equation then gives for the higher order coefficients 
\begin{eqnarray} 
   a_1 &=& - \frac{\sqrt{3}}{36}\Clpt 
             + \frac{1}{24} = -0.00716473... \ , \nonumber\\ 
   a_2 &=& \sqrt{3} \frac{5}{324}\Clpt 
           - \frac{23}{864} = 0.000508185... \ , \nonumber\\ 
   a_3 &=& - \sqrt{3} \frac{31}{2916}\Clpt 
           + \frac{145}{7776} = -0.0000414414...\ , 
\label{a1a3} 
\end{eqnarray} 
and so on. The rational fractions (such as $1/24$ in $a_1$) are the 
expansion of the solution of the complete inhomogenous equation 
Eq.(\ref{diffeq}) with initial condition $S(2,0)=0$, whereas the 
irrational part with $\Clpt$, which is proportional to the regular solution of the associated homogenous equation, is fixed by the initial conditions. 
\par 
The nearest singular point of the differential equation Eq.(\ref{diffeq}) 
is at $z=-1$ ($s=1$), so one would expect a convergence radius 1 for 
the expansion Eq.(\ref{expat0}); but the nearest singularity of $S(2,z)$ 
is at the physical threshold $s=9$ or $z=-9$, so that $S(2,z)$ is 
analytic at 
the  pseudothreshold
$z=-1$; that is achieved by suitable cancellations among the two 
(rational and irrational)
terms which appear in the explicit expressions of the coefficients 
$a_n$, as already shown also by the first coefficients Eq.(\ref{a1a3}). 
\par 
The next singular point of the differential equation is at $z=-9$ (or 
$s=9$), which is also a non analyticity point for $S(2,z)$, as 
commented above; the expansion Eq.(\ref{expat0}) has therefore convergence 
radius 9. 
\par 
The knowledge of the position of the nearest singularity allows to 
greatly speed up the convergence of the series, by introducing 
the ``Bernoulli" variable (first proposed in \cite{'tHooft:1978xw} for the 
evaluation of Euler's dilogarithm, and repeatedly used in 
\cite{Gehrmann:2001pz} for the evaluation of Harmonic Polylogarithms),  
which is given in this case by 
\begin{equation} 
      t = \ln \left(1 + \frac{z}{9} \right) . 
\label{ztat0} 
\end{equation} 
One has 
$$ z=9(\e^t-1) = 9\left(t +\frac{t^2}{2} + \frac{t^3}{6}+\dots\right) \ ; $$ 
replacing $z$ in Eq.(\ref{expat0}) by its expansion in $t$, 
one obtains a new expansion in $t$ which we write as 
\begin{equation} 
     S(2,z) = \sum_{n=0}^\infty \alpha_n t^n \ . 
\label{expat0int} 
\end{equation} 
The explicit knowledge of $N$ terms of the expansion Eq.(\ref{expat0}) 
detrmines the first $N$ terms of Eq.(\ref{expat0int}), 
which for a same number of expansion terms and small $t$ (see 
Section 6 for more details) 
approximates $S(2,z)$ much better than the original expansion in $z$. 
A straightforward calculation gives for the first coefficients 
\begin{eqnarray} 
   \alpha_0 &=& a_0 = \frac{\sqrt{3}}{12}\Clpt=0.146494... \ ,         \nonumber\\ 
   \alpha_1 &=& \frac{3}{8} - \frac{\sqrt{3}}{4}\Clpt=-0.0644826... \ ,  \nonumber\\ 
   \alpha_2 &=& \sqrt{3} \frac{9}{8}\Clpt - \frac{63}{32} 
                                         = 0.00892173... \ , \nonumber\\ 
   \alpha_3 &=& - \sqrt{3} \frac{157}{24}\Clpt 
                  + \frac{23}{2} = 0.000205110... \ , 
\label{alpha03} 
\end{eqnarray} 
and so on. Note the strong cancellation between the two terms in each 
coefficient of the expansion; the cancellation, already present in 
Eq.(\ref{a1a3}) as remarked above, is even bigger in Eq.(\ref{alpha03}), 
and grows quickly with the order $n$ of the coefficient $\alpha_n$. 
Keeping the control of cancellations in the numerical evaluation of 
the coefficients would be in principle a delicate task, but that is 
no longer problem when using the arbitrary precision features of 
{\tt Mathematica} \cite{mathematica}.

\section{ The expansion around $z=-9$. } 
As already remarked, the point $z=-9$ ($s=9$) is the nearest singular point 
of $S(2,z)$. For $z\to-9$, the general solution consists of a regular 
part and a singular part behaving like $\ln(z+9)$, so that 
$S(2,z)$ admits an expansion of the general form 
\begin{equation} 
     S(2,z) = \sum_{n=0}^\infty a_n (z+9)^n 
            + \ln(z+9) \sum_{n=0}^\infty b_n (z+9) ^n \ , 
\label{Sexpat9} 
\end{equation} 
where the coefficients, called again $a_n, b_n$ as in 
Eq.(\ref{Gexpat0}), are of course 
different from the coefficients 
of the expansion at $z=0$. 
For $z >-9 $, {\it i.e.} $s<9$ or $s$ below threshold, both terms 
of Eq.(\ref{Sexpat9}) are real, while for $z<-9$, or $s>9$, above 
threshold, $\ln(z+9)$ develops an immaginary part, which with the 
usual $s+i\epsilon$ prescription is $-i\pi$, 
\begin{equation} 
   \ln(z+9) = \ln|z+9| - i\ \pi \theta(-z-9) \ . 
\label{ln(z+9)} 
\end{equation} 
\par 
From \cite{Laporta:2004rb} we have the initial conditions 
\begin{eqnarray} 
   a_0 &=& - \frac{\sqrt{3}}{48} \pi \ ,        \nonumber\\ 
   b_0 &=& \frac{\sqrt{3}}{48} \left[\pi \ln(72) - 5\Clpt \right] \ . 
\label{z9init} 
\end{eqnarray} 
From the differential equation we found for the first coefficients 
\begin{eqnarray} 
     a_1 &=& a_0 \frac{1}{12} + b_0 \frac{5}{72} + \frac{1}{192} 
                                               \ , \nonumber\\ 
     b_1 &=& b_0 \frac{1}{12} \ ,                  \nonumber\\ 
     a_2 &=& a_0 \frac{7}{864} + b_0 \frac{97}{10368} + \frac{5}{6912} 
                                               \ , \nonumber\\ 
     b_2 &=& b_0 \frac{7}{864} \ , 
\label{ab1ab2} 
\end{eqnarray} 
and so on for the higher order coefficients. As in the previous section, 
the terms independent of the initial conditions $a_0, b_0$ (such as 
the fraction $1/192$ in $a_1$) are due to the inhomogenous term in the 
differential equation Eq.(\ref{diffeq}). 
\par 
The nearest singular point for the two components of the expansion 
Eq.(\ref{Sexpat9}) is at $z=-1$, so that the radius of convergence of the 
two expansions is $8$, {\it i.e.} the expansions converge in the 
range $ -17 < z < -1 $ or $ 1 < s < 17 $. 
That might look surprising, as 
$S(2,z)$ is analytic at $z=-1$; but the two terms in the expansion 
Eq.(\ref{Sexpat9}) cannot combine to a single expansion converging in a 
wider range, given the presence of the $\ln(z+9)$ in front of the 
second term. \par 
In this case the ``Bernoulli" variable is therefore 
\begin{equation} 
      t = - \ln \left(1 - \frac{z+9}{8} \right) , 
\label{ztat9} 
\end{equation} 
and the expansion of $S(2,z)$ in terms of $t$ reads 
\begin{equation} 
     S(2,z) = \sum_{n=0}^\infty \alpha_n t^n 
            + \ln(z+9) \sum_{n=0}^\infty \beta_n t^n \ . 
\label{expat9int} 
\end{equation} 
The explicit calculation gives 
\begin{eqnarray} 
     \alpha_0 &=& a_0=0.301695... \ ,  \nonumber\\ 
     \beta_0  &=& b_0=-0.113362... \ ,  \nonumber\\ 
     \alpha_1 &=& \frac{2}{3} a_0 + \frac{5}{9} b_0  + \frac{1}{24}
               =0.179817...  \ , \nonumber\\ 
     \beta_1  &=& \frac{2}{3} b_0=-0.0755749...  \ , \nonumber\\ 
     \alpha_2 &=& \frac{5}{27} a_0 + \frac{26}{81} b_0  + \frac{11}{432} 
               =0.0449445... \ , \nonumber\\ 
     \beta_2  &=& \frac{5}{27} b_0=-0.0209930...  \ , 
\label{alphab1ab2} 
\end{eqnarray} 
with the by now usual strong cancellations within the various terms 
of each coefficient. 

\section{ The expansion at infinity. } 
The differential equation has also $z=\infty$ as singular point; for 
$z \to \infty $ the expected behaviour for $S(2,z)$ is 
\begin{equation}
     S(2,z) = \frac{1}{z}\left( \sum_{n=0}^\infty  a_n \frac{1}{z^n} 
            + \ln{z} \sum_{n=0}^\infty b_n \frac{1}{z^n} 
            + \ln^2{z} \sum_{n=0}^\infty c_n \frac{1}{z^n} \right) \ . 
\label{Sexpatoo} 
\end{equation} 
For $z>0$ ({\it i.e.} $s$ spacelike) the previous expression is real; 
in the timelike region, we have to replace $z$ by  $z-i\epsilon$; 
that gives 
\begin{eqnarray} 
         \ln{z}   &=& \ln|z| - i\pi\theta(-z) \ , \nonumber\\ 
         \ln^2{z} &=& \ln^2|z| 
                   - \theta(-z) \left( 2i\pi\ln|z| + \pi^2 \right) 
                                            \ , 
\label{lnz} 
\end{eqnarray} 
from which one can obtain separately the real and immaginary parts 
of $S(2,z)$. 
\par 
From \cite{Laporta:2004rb} the initial conditions are 
\begin{eqnarray} 
     a_0 = 0 \ , \nonumber\\ 
     b_0 = 0 \ , \nonumber\\ 
     c_0 = \frac{3}{16} \ . 
\label{zooinit} 
\end{eqnarray} 
It is to be noted that the presence of the term $ 3/16\ln^2{z} $ is 
anyhow required by the inhomogenous term in the differential equation, 
so that strictly speaking the initial conditions are simply 
$a _0 = b_0 = 0$. The differential equation then gives for the first 
coefficients 
\begin{eqnarray} 
     a_1 &=& \frac{3}{8}   \ , \nonumber\\ 
     b_1 &=& \frac{3}{2}    \ , \nonumber\\ 
     c_1 &=& - \frac{9}{16}   \ , \nonumber\\ 
     a_2 &=& - \frac{3}{32}   \ , \nonumber\\ 
     b_2 &=& - \frac{39}{4}    \ , \nonumber\\ 
     c_2 &=& \frac{45}{16}   \ , 
\label{abc1abc2} 
\end{eqnarray} 
and so on for the higher orders. 
\par 
As the nearest singularity is at $z=-9$, the ``Bernoulli" variable is 
\begin{equation} 
      t = \ln \left(1 + \frac{9}{z} \right) , 
\label{ztatoo} 
\end{equation} 
and we write the expansion of $S(2,z)$ in terms of $t$ as 
\begin{equation}
     S(2,z) = \left( \sum_{n=0}^\infty  \alpha_n t^n 
            + \ln{z} \sum_{n=0}^\infty \beta_n t^n 
            + \ln^2{z} \sum_{n=0}^\infty \gamma_n t^n \right) \ , 
\label{expatooint} 
\end{equation} 
so that the first coefficients are 
\begin{eqnarray} 
     \alpha_0 &=& \beta_0 = \gamma_0 = 0 \ , \nonumber\\ 
     \alpha_1 &=& a_0 = 0   \ , \nonumber\\ 
     \beta_1 &=& b_0 = 0 \ , \nonumber\\ 
     \gamma_1 &=& \frac{1}{48}   \ , \nonumber\\ 
     \alpha_2 &=& \frac{1}{216}   \ , \nonumber\\ 
     \beta_2 &=& \frac{1}{54}   \ , \nonumber\\ 
     \gamma_2 &=& \frac{1}{288}   \ , \nonumber\\ 
     \alpha_3 &=& \frac{35}{776}   \ , \nonumber\\ 
     \beta_3 &=& \frac{5}{972}   \ , \nonumber\\ 
     \gamma_3 &=& \frac{1}{2592}   \ , 
\label{alphabc1abc2} 
\end{eqnarray} 
etc.

\section{ The {\tt FORTRAN} numerical routine. } 
As a first step, we obtained the analytic expression of the coefficients 
of all the needed expansions for $S(2,z)$ up to the required order 
(see below). From those we obtained the corresponding expansions 
for the other three quantities, $S_1(2,z), S^{(0)}(4,z) $ and 
$ S_1^{(0)}(4,z) $ by means of the formulae 
Eq.s(\ref{S1(2)},\ref{S(0,4)},\ref{S1(0,4)}), 
which involve only elementary algebra and differentiation. 
When dealing with power series expansions, differentiating is trivial, 
even if one term is lost at each differentiation; but that is easily 
compensated by allowing for a few extra terms in the initial expansion. 
\par 
We then evaluated the numerical value of all the coefficients so obtained. 
The analytic and numerical evaluations were carried out by 
{\tt Mathematica}, in arbitrary precision arithmetics mode for keeping 
rounding errors under check. The numerical values of the coefficients 
were transferred to the {\tt FORTRAN} code as constants with $18$ 
digits; that exceeds of the course the double precision of {\tt FORTRAN} 
but eliminates rounding problems in the least significant digit. 
\par 
In the range $ -\infty < z < -17.45 $, or $ 17.45 < s < +\infty$ 
(timelike region above threshold, all MIs complex), we evaluate the 4 
functions by means of truncated expansions of the form 
\begin{equation} 
       \sum_{n=0}^N \alpha_n t^n 
     + \ln{z} \sum_{n=0}^N\beta_n t^n 
     + \ln^2{z} \sum_{n=0}^N\gamma_n t^n \ , 
\label{-oo-17.45} 
\end{equation} 
where $t$ is defined by Eq.(\ref{ztatoo}) and varies therefore in the range 
$  -0.725173... < t < 0 $, the complex value of $\ln{z}$ is specified in 
Eq.(\ref{lnz}) and the values of the coefficients $\alpha_n,\beta_n,\gamma_n$
depend of course  on the MIs.
 We obtain the required double precision  
(relative error less than $1\times 10^{-15}$) for all the 4 
functions with $N=22$. 
Those numerical checks, beyond the possibilities offered by double 
precision {\tt FORTRAN}, were carried out by using {\tt Mathematica}
in arbitrary precision mode and including a growing number of terms in the 
expansion. 
\par 
In the range $ - 17.45 \le z < -5.15 $, or  $ 5.15 < s \le 17.45$ 
(timelike region across the threshold $s=9$) 
we use for the 4 functions truncated expansions of the form 
\begin{equation} 
              \frac{R}{z+9} 
            + \sum_{n=0}^N \alpha_n t^n 
            + \ln(z+9) \sum_{n=0}^N \beta_n t^n \ , 
\label{-17.45-5.15} 
\end{equation} 
where $R=-\sqrt{3}\,\pi/48$ for $S_1(2,z)$ and vanishes in all the other cases, 
$t$, defined in Eq.(\ref{ztat9}), varies in the range 
$ -0.720884... \le t < 0.656333... $ and 
the value of $\ln(z+9)$ is given in 
Eq.(\ref{ln(z+9)}). The required precision (better than $1\times 10^{-15}$) is obtained with $N=21$. 
\par 
In the range $ -5.15 \le z < 11 $, 
or $ -11 < s \le 5.15 $, we use truncated expansions of the form  
\begin{equation} 
     \sum_{n=0}^N \alpha_n t^n, 
\label{-5.15+11} 
\end{equation} 
where $t$, defined in Eq.(\ref{ztat0}), varies in the range 
$ -0.849151 \le t < 0.798508 $, and the required precision is obtained 
with $N=18$. Note that the expansion in $t$ is valid, and still 
quickly convergent, for corresponding values of $z$ also in the region 
$ 9 < z < 11 $, 
which lies outside the region of convergence of
the expansion in $z$ from which the expansion in $t$ was derived! 
\par 
Finally, in the range $ 11 \le z < +\infty $, or $-\infty < s \le -11$ 
(spacelike region), we use again the expansion Eq.(\ref{-oo-17.45}), 
with $t$ varying in the range $ 0 < t \le  0.597837... $; in this case 
$\ln{z}$ is real, and the aimed precision is obtained with $N=20$. 
\par 
The numerical values of the 4 functions, evaluated according to the 
previous formulas, are returned by calling the 
{\tt FORTRAN} routine \\ 
{\centerline  
{ \tt SUBROUTINE SUNRISE(S,SR2,SI2,S1R2,S1I2,SR4,SI4,S1R4,S1I4) \ , } } 
where all the variables are real {\tt DOUBLE PRECISION} variables, 
{\tt S} is the momentum transfer ({\tt S}, equal 
to $-z$ of previous sections, is positive when timelike, with physical 
threshold at $9$),
{\tt SR2,SI2} are the real and immaginary parts of $S(2,z)$ as defined 
in Eq.(\ref{defMIs}) for $d=2$ and $z=-{\tt S}$ , 
{\tt S1R2,S1I2} the real and immaginary parts of $S_1(2,z)$ as defined 
also in Eq.(\ref{defMIs}), 
{\tt SR4,SI4} the real and immaginary parts of $S^{(0)}(4,z)$ as defined 
by Eq.s(\ref{defMIs},\ref{expS}), and {\tt S1R4,S1I4} finally 
the real and immaginary parts of $S_1^{(0)}(4,z)$. 
{\tt S} can vary in the whole real axis, $ -\infty < {\tt S} < +\infty $; 
the immaginary parts, which appear when {\tt S} is above threshold 
(${\tt S}>9$), are evaluated according to the standard 
${\tt S}+i\epsilon$ prescription. 
\par 
The calculation is carried out with relative 
precision better than $1\times 10^{-15}$ (which is essentialy the double precision limit of {\tt FORTRAN}), except of course when the corresponding quantities 
approach zero.
$S(2,z)$ and $S_1(2,z)$ are positive definite and tend to zero for $|z|\to \infty$, while 
$S^{(0)}(4,z)$ and $S_1^{(0)}(4,z)$ have a zero 
around $z=12.5910...$ and $z=-3.5599...$, respectively.
\par 
Internally, the routine {\tt SUNRISE} calls, depending on the actual 
value of {\tt S}, one of the three subroutines 
{\tt SUNRISE0, SUNRISE9} and {\tt SUNRISEOO}. Those routines implement 
the various expansions as discussed in the previous 
sections. For ease of check and implementation, the routines contain 
more coefficients than actually used, each coefficient being an entry in a 
{\tt DATA} statement written as a constant with $18$ decimal digits (as 
already remarked above, that 
exceeds of course the double precision of {\tt FORTRAN} but eliminates 
rounding problems). The arguments of the three routines 
consist of the same arguments of {\tt SUNRISE} and an extra integer 
argument, specifying how many coefficients of the expansions are to be 
considered in the calculation. 
The three internal routines can in principle be called independently, 
for studying the contributions of the various terms of the expansions 
etc., as well as for additional checks. 
\par 
As a last remark, we succeeded in covering the whole real 
axis in $z$ with just three different expansions and a maximum of 
$22$ terms at worst; that is fully satisfactory, and from the practical 
point of view there is no need of further improvements. 
But let us observe that, in principle, the calculation could be further 
speeded up by using additional auxiliary expansions. For instance, one could 
consider another expansion point, say for instance at $z=-18$ 
(or $s=18$); as that is a regular point in which the function is 
analytic, one can write as 
$$ A + B (z+18) +\ord\left((z+18)^2\right)$$ 
the limiting behaviour of the function to be evaluated, parametrizing it 
in terms of two initial constants $A,B$, still to be determined.  
By using the differential equation, it is immediate to generate as many 
coefficients as desired of the expansion in powers of $(z+18)$. 
The two constants can then be obtained 
by matching numerically the new expansion with the expansion around 
$z=-9$ at, say, $z=-13$, and with the expansion at $z=-\infty$ at, 
say, $z = 24$. In so doing, one has to use every involved expansion on a 
smaller range, where the desired precision can be achieved with a smaller 
number of terms. 
\par 
By the same token, one could consider the expansion around $z=-9$ 
Eq.(\ref{Sexpat9}) and keep the initial conditions $a_0, b_0$, 
whose exact value is given by Eq.(\ref{z9init}), as still unknown 
constants, use the differential equation to work out the desired 
coefficients of the expansion around $z=-9$, and then obtain 
numerically $a_0, b_0$ by imposing the smooth matching at, say, 
$z=-5$ with the value obtained with the expansion around $z=0$, 
supposedly known. One could then continue to recover in that manner the 
numerical value of the initial constants of the expansion at infinity, 
supposedly unknown as well. Note that in this procedure one can 
insert as many auxiliary expansions as considered useful; but one cannot 
skip the expansions at the singular points of the function, 
as those points cannot be included, by the very definition 
of singular points, in the convergence radius of any expansion 
around nearby points. The differential equation, again, gives the 
most general behaviour of the function at any point, to be used 
for obtaining the expansion around that point. 

\section{Example}
The following simple program illustrates how to evaluate the
real and imaginary parts of the Master integrals 
$S(2,z)$, $S_1(2,z)$, $S^{(0)}(4,z)$ and $S_1^{(0)}(4,z)$
for a given value of the momentum transfer $s=-z$:
\begin{verbatim}
      program examplesunem
      implicit none
      real*8 S,SR2,SI2,S1R2,S1I2,SR4,SI4,S1R4,S1I4

      write(6,*)'Input squared momentum transfer S:'
      read(5,*) S

      call sunrise(S,SR2,SI2,S1R2,S1I2,SR4,SI4,S1R4,S1I4) 

      write(6,*)'First Master integral in d=2 dimensions'
      write(6,200) S,SR2
      write(6,201) S,SI2
      write(6,*)'Second Master integral in d=2 dimensions'
      write(6,210) S,S1R2
      write(6,211) S,S1I2
      write(6,*)'First Master integral in d=4 dimensions: finite parts'
      write(6,400) S,SR4
      write(6,401) S,SI4
      write(6,*)'Second Master integral in d=4 dimensions: finite parts'
      write(6,410) S,S1R4
      write(6,411) S,S1I4

 200  format('  Re S(2,',1PE22.15,')     = ',1PE22.15)
 201  format('  Im S(2,',1PE22.15,')     = ',1PE22.15)
 210  format('  Re S1(2,',1PE22.15,')    = ',1PE22.15)
 211  format('  Im S1(2,',1PE22.15,')    = ',1PE22.15)

 400  format('  Re S^(0)(4,',1PE22.15,') = ',1PE22.15)
 401  format('  Im S^(0)(4,',1PE22.15,') = ',1PE22.15)
 410  format('  Re S1^(0)(4,',1PE22.15,')= ',1PE22.15)
 411  format('  Im S1^(0)(4,',1PE22.15,')= ',1PE22.15)
 
      end
\end{verbatim}

\section*{Acknowledgements}

This work was supported in part by the Deutsche Forschungsgemeinschaft 
in the SFB/TR 09-03.

 

\begin{thebibliography}{99}

\bibitem{Laporta:2004rb} 
S.~Laporta and E.~Remiddi, 
Nucl.\ Phys.\ B {\bf 704} (2005) 349 
[hep-ph/0406160]. 

\bibitem{Remiddi:1999ew}
E.~Remiddi and J.~A.~M.~Vermaseren,
Int.\ J.\ Mod.\ Phys.\ A {\bf 15} (2000) 725
[hep-ph/9905237].

\bibitem{'tHooft:1978xw} 
G.~'t Hooft and M.~J.~G.~Veltman, 
Nucl.\ Phys.\ B {\bf 153} (1979) 365. 

\bibitem{Gehrmann:2001pz} 
T.~Gehrmann and E.~Remiddi, 
Comput.\ Phys.\ Commun.\  {\bf 141} (2001) 296 
[hep-ph/0107173]. 


\bibitem{mathematica}
S.~Wolfram, Mathematica Version 4.2.

\end{thebibliography}
\end{document}